\documentclass[aps,preprint]{revtex4}%
\usepackage{amsfonts}
\usepackage{amsmath}
\usepackage{amssymb}
\usepackage{subfigure}
\usepackage{graphicx}%
\begin{document}
\preprint{CTP-SCU/2017006}
\title{Action Growth in $f\left(  R\right)  $ Gravity}
\author{Peng Wang}
\email{pengw@scu.edu.cn }
\author{Haitang Yang}
\email{hyanga@scu.edu.cn}
\author{Shuxuan Ying}
\email{2013222020007@stu.scu.edu.cn}
\affiliation{Center for Theoretical Physics, College of Physical Science and Technology,
Sichuan University, Chengdu, 610064, China}

\begin{abstract}
Inspired by the recent \textquotedblleft Complexity = Action\textquotedblright%
\ conjecture, we use the approach proposed by Lehner \textit{et al}. to
calculate the rate of the action of the WheelerDeWitt patch at late times for
static uncharged and charged black holes in $f\left(  R\right)  $ gravity. Our
results have the same expressions in terms of the mass, charge, and electrical
potentials at the horizons of black holes as in Einstein's gravity. In the
context of $f\left(  R\right)  $ gravity, the Lloyd bound is saturated for
uncharged black holes but violated for charged black holes near extremality.
For charged black holes far away from the ground states, the Lloyd bound is
violated in four dimensions but satisfied in higher dimensions.

\end{abstract}
\maketitle
\tableofcontents



\section{Introduction}

Recently, Brown \textit{et al. }\cite{IN-Brown:2015bva,IN-Brown:2015lvg}
proposed the \textquotedblleft Complexity = Action\textquotedblright%
\ (CA)\ duality, which conjectures that the computational complexity
$\mathcal{C}$ of a holographic boundary state could be identified with the
classical gravitational action $S_{\text{WdW}}$ of the WheelerDeWitt patch:%
\begin{equation}
\mathcal{C=}\frac{S_{\text{WdW}}}{\pi\hbar}. \label{eq:CA}%
\end{equation}
The WheelerDeWitt patch is defined as the domain of dependence of any Cauchy
surface anchored at the boundary state. Loosely speaking, the complexity
$\mathcal{C}$ of a state is the minimum number of quantum gates to prepare
this state from a reference state \cite{IN-Wat:2009,IN-Huang:2015,IN-Osb:2011}%
. The CA\ duality is the refined version of the \textquotedblleft Complexity =
Volume\textquotedblright\ duality
\cite{IN-Susskind:2014rva,IN-Stanford:2014jda,IN-Susskind:2014jwa,IN-Susskind:2014moa}%
, which states that the complexity of a boundary state is dual to the volume
of the maximal spatial slice crossing the Einstein-Rosen bridge anchored at
the boundary state. Later, the \textquotedblleft Complexity = Volume
2.0\textquotedblright\ duality was proposed in \cite{IN-Couch:2016exn}, in
which the complexity was identified with the spacetime volume of the
WheelerDeWitt patch.

After calculating the action growth $dS_{\text{WdW}}/dt$ for various
stationary AdS black holes in \cite{IN-Brown:2015lvg}, Lehner \textit{et al}.
\cite{IN-Lehner:2016vdi} carefully analyzed the action of some subregion with
null segments and joints at which a null segment was joined to another
segment. A set of rules for calculating the contributions from these joints
were also given in \cite{IN-Lehner:2016vdi}. Although the two approaches in
\cite{IN-Brown:2015lvg} and \cite{IN-Lehner:2016vdi} look quite different,
they gave the same results for various black holes within Einstein's gravity
\cite{IN-Lehner:2016vdi}. Beyond Einstein's gravity, the action growth was
calculated by the method of \cite{IN-Brown:2015lvg} in cases of Gauss-Bonnet
gravity \cite{IN-Cai:2016xho}, massive gravities \cite{IN-Pan:2016ecg},
$f\left(  R\right)  $ gravity \cite{IN-Alishahiha:2017hwg}, and critical
gravities \cite{IN-Alishahiha:2017hwg}. On the other hand, following the
method of \cite{IN-Lehner:2016vdi}, the action growth was calculated for
Born-Infeld black holes \cite{IN-Cai:2017sjv,IN-Tao:2017fsy}, charged dilaton
black holes \cite{IN-Cai:2017sjv}, and charged black holes with phantom
Maxwell field \cite{IN-Cai:2017sjv} in AdS space. Moreover, the divergent
terms of $S_{\text{WdW}}$ due to the infinite volume near the boundary of AdS
space were considered in
\cite{IN-Carmi:2016wjl,IN-Reynolds:2016rvl,IN-Kim:2017lrw}, where it showed
that these terms could be written as local integrals of boundary geometry.

One of the simplest modifications to Einstein's gravity is the $f\left(
R\right)  $ gravity
\cite{IN-Bergmann:1968ve,IN-Capozziello:2009nq,IN-Capozziello:2011et,IN-Nojiri:2010wj}
in which the Lagrangian density $f$ is an arbitrary function of $R$, where $R$
is the Ricci scalar. It can be shown that the metric-$f\left(  R\right)  $
gravity is equivalent to the $\omega_{\text{BD}}$ $=0$ Brans-Dicke theory with
the potential \cite{IN-DeFelice:2010aj}. In \cite{IN-Alishahiha:2017hwg}, the
action growth for static uncharged black holes in $f\left(  R\right)  $
gravity was calculated using the method of \cite{IN-Brown:2015lvg}. It is
interesting to calculate the action growth in $f\left(  R\right)  $ gravity
using the method of \cite{IN-Lehner:2016vdi} and then check whether these two
results are same. In this paper, we will employ the approach proposed in
\cite{IN-Lehner:2016vdi} to compute $dS_{\text{WdW}}/dt$ at late times for
static uncharged and charged black holes in $f\left(  R\right)  $ gravity.

The remainder of our paper is organized as follows: In section \ref{Sec:AfG},
we discuss the boundary terms in the action functional of $f\left(  R\right)
$ gravity when the boundary includes null segments. In order to employ the
method of \cite{IN-Lehner:2016vdi}, we consider the Einstein frame
representation of the action of a Brans-Dicke theory with Brans-Dicke
parameter $\omega_{\text{BD}}=0$, which is dynamically equivalent to the
metric-$f\left(  R\right)  $ gravity. In section \ref{Sec:AGBHfG}, the action
growth of the WheelerDeWitt patch is calculated in the cases of static
uncharged and charged black holes in $f\left(  R\right)  $ gravity. In section
\ref{Sec:Con}, we conclude with a brief discussion of our results.

\section{Action in $f\left(  R\right)  $ Gravity}

\label{Sec:AfG}

The action that defines $f\left(  R\right)  $ gravity has the generic form%
\begin{equation}
S=\int d^{d+1}x\sqrt{-g}f\left(  R\right)  +S^{m}\left(  g_{\mu\nu}%
,\psi\right)  \text{,}\label{eq:action}%
\end{equation}
where $S^{m}$ is the matter action, $\psi$ is the matter field, and we take
$16\pi G=1$. The gravitational equation can be derived by varying the action
$\left(  \ref{eq:action}\right)  $ with respect to $g_{\mu\nu}$:%
\begin{equation}
f^{\prime}\left(  R\right)  R_{\mu\nu}-\frac{1}{2}f\left(  R\right)  g_{\mu
\nu}+\left(  g_{\mu\nu}\nabla^{2}-\nabla_{\mu}\nabla_{\nu}\right)  f^{\prime
}\left(  R\right)  =\frac{1}{2}T_{\mu\nu}^{m},\label{eq;EOM}%
\end{equation}
where $T_{\mu\nu}^{m}$ is the energy-momentum tensor of the matter field
defined by%
\begin{equation}
T_{\mu\nu}^{m}=-\frac{2}{\sqrt{-g}}\frac{\delta S^{m}}{\delta g^{\mu\nu}}.
\end{equation}
Introducing a new field $\chi$, we could rewrite the action $\left(
\ref{eq:action}\right)  $ as a dynamically equivalent action:%
\begin{equation}
S=\int d^{d+1}x\sqrt{-g}\left[  f^{\prime}\left(  \chi\right)  \left(
R-\chi\right)  +f\left(  \chi\right)  \right]  +S^{m}\left(  g_{\mu\nu}%
,\psi\right)  .\label{eq:actionx}%
\end{equation}
Varying the action $\left(  \ref{eq:actionx}\right)  $ with respect to $\chi$
gives%
\begin{equation}
f^{\prime\prime}\left(  \chi\right)  \left(  R-\chi\right)  =0.
\end{equation}
Therefore, $\chi=R$ if $f^{\prime\prime}\left(  \chi\right)  \neq0$, which
reproduces the action $\left(  \ref{eq:action}\right)  $. With $\chi=R$, the
equation of motion (EOM) obtained by vary the action $\left(  \ref{eq:actionx}%
\right)  $ with respect to $g_{\mu\nu}$ recovers eqn. $\left(  \ref{eq;EOM}%
\right)  $. Redefining $f^{\prime}\left(  \chi\right)  $ as a field, it shows
that the action $\left(  \ref{eq:actionx}\right)  $ is the Jordan frame
representation of the action of a Brans-Dicke theory with Brans-Dicke
parameter $\omega_{\text{BD}}=0$. To diagonalizes the gravi-$\chi$ kinetic
term, we introduce the rescaled metric $\tilde{g}$ in the $\tilde{x}$
coordinate:%
\begin{equation}
\tilde{g}_{\tilde{\mu}\tilde{\nu}}d\tilde{x}^{\tilde{\mu}}d\tilde{x}%
^{\tilde{\nu}}=f^{\prime}\left(  \chi\right)  ^{\frac{2}{d-1}}g_{\mu\nu
}dx^{\mu}dx^{\nu}.\label{eq:rescale}%
\end{equation}
The action $\left(  \ref{eq:actionx}\right)  $ then becomes%
\begin{align}
S &  =\int d^{d+1}\tilde{x}\sqrt{-\tilde{g}}\tilde{R}+\frac{\sqrt{2d}}%
{\sqrt{d-1}}\int d^{d+1}\tilde{x}\sqrt{-\tilde{g}}\tilde{\nabla}^{2}%
\phi\nonumber\\
&  -\frac{1}{2}\int d^{d+1}\tilde{x}\sqrt{-\tilde{g}}\left\{  \left(
\tilde{\nabla}\phi\right)  ^{2}+2f^{\prime}\left(  \chi\right)  ^{-\frac
{d+1}{d-1}}\left[  f^{\prime}\left(  \chi\right)  \chi-f\left(  \chi\right)
\right]  \right\}  +S^{m}\left[  f^{\prime}\left(  \chi\right)  ^{-\frac
{2}{d-1}}\tilde{g}_{\tilde{\mu}\tilde{\nu}},\psi\right]  ,\label{eq:actionE}%
\end{align}
where
\[
\phi=\sqrt{\frac{2d}{d-1}}\ln f^{\prime}\left(  \chi\right)  .
\]

Now consider the action $\left(  \ref{eq:actionE}\right)  $ over a region
$\mathcal{V}$ of spacetime with the boundary $\partial\mathcal{V}$. Since
there are second derivatives of the metric tensor and the field $\phi$ in the
first line of eqn. $\left(  \ref{eq:actionE}\right)  $, extra boundary terms
need to be added to derive the EOMs from the action. The $\tilde{\nabla}%
^{2}\phi$ term in eqn. $\left(  \ref{eq:actionE}\right)  $ can be expressed as
a boundary term via Stokes's theorem:%
\begin{equation}
\int_{\mathcal{V}}d^{d+1}\tilde{x}\sqrt{-\tilde{g}}\tilde{\nabla}^{2}\phi
=\int_{\partial\mathcal{V}}d^{d}\tilde{x}\sqrt{\left\vert \tilde{h}\right\vert
}n^{\tilde{\mu}}\tilde{\nabla}_{\tilde{\mu}}\phi,
\end{equation}
where $\tilde{h}_{\tilde{\mu}\tilde{\nu}}$ is the induced metric on
$\partial\mathcal{V}$, and $n^{\tilde{\mu}}$ is the unit vector normal to
$\partial\mathcal{V}$. To have the EOMs by variation of action, this boundary
term should be canceled against by another one%
\begin{equation}
S_{\partial\mathcal{V}}^{\phi}=-\frac{\sqrt{2d}}{\sqrt{d-1}}\int
_{\partial\mathcal{V}}d^{d}\tilde{x}\sqrt{\left\vert \tilde{h}\right\vert
}n^{\tilde{\mu}}\tilde{\nabla}_{\tilde{\mu}}\phi. \label{eq:Bphi}%
\end{equation}
The first term in eqn. $\left(  \ref{eq:actionE}\right)  $ is just the
standard Hilbert action in terms of $\tilde{g}_{\tilde{\mu}\tilde{\nu}}$,
which contains second derivatives of $\tilde{g}_{\tilde{\mu}\tilde{\nu}}$ and
hence requires extra boundary terms to cancel against boundary contributions
from $\tilde{R}$ to find the EOMs. These extra boundary terms were carefully
discussed in \cite{IN-Lehner:2016vdi}. The terms in the second line of eqn.
$\left(  \ref{eq:actionE}\right)  $ contain at most first derivative of fields
and do not need extra boundary terms to obtain the EOMs. Following conventions
in \cite{IN-Carmi:2016wjl}, the action over the region $\mathcal{V}$ including
boundary terms is given by%
\begin{equation}
S=S_{\mathcal{V}}+S_{\partial\mathcal{V}}^{\phi}+S_{\partial\mathcal{V}}%
^{g}\text{,} \label{eq:TotalS}%
\end{equation}
where $S_{\mathcal{V}}$ is $S$ given by eqn. $\left(  \ref{eq:actionE}\right)
$ evaluated over $\mathcal{V}$, $S_{\partial\mathcal{V}}^{\phi}$ is given by
eqn. $\left(  \ref{eq:Bphi}\right)  $, and
\begin{equation}
S_{\partial\mathcal{V}}^{g}=2\int_{\mathcal{B}}d^{d}\tilde{x}\sqrt{\left\vert
\tilde{h}\right\vert }K-2\int_{\mathcal{B}^{\prime}}d\lambda d^{d-1}%
\theta\sqrt{\gamma}\kappa+2\int_{\Sigma}d^{d-1}\tilde{x}\sqrt{\tilde{\sigma}%
}\eta+2\int_{\Sigma^{\prime}}d^{d-1}\tilde{x}\sqrt{\tilde{\sigma}}a.
\label{eq:Bg}%
\end{equation}
In $\left(  \ref{eq:Bg}\right)  $, $\mathcal{B}$ denotes the spacelike or
timelike segments of $\partial\mathcal{V}$ while $\mathcal{B}^{\prime}$
denotes the null segments. The $\Sigma^{\prime}$ denotes joints involving null
boundaries, and $\Sigma$ denotes other joints. The definitions of other
quantities can be found in \cite{IN-Carmi:2016wjl}. It is noteworthy that we
could choose an affine parametrization for each null surface, and these make
no contribution to the action. When the fields satisfy the EOM, the values of
the actions $\left(  \ref{eq:action}\right)  $ and $\left(  \ref{eq:actionE}%
\right)  $ are same. In this case, one could have%
\begin{equation}
S_{\mathcal{V}}=\int_{\mathcal{V}}d^{d+1}x\sqrt{-g}f\left(  R\right)
+S_{\mathcal{V}}^{m}\left(  g_{\mu\nu},\psi\right)  , \label{eq:Sf(R)}%
\end{equation}
where $S_{\mathcal{V}}^{m}\left(  g_{\mu\nu},\psi\right)  $ is the matter
action evaluated over $\mathcal{V}$.

\section{Action Growth of Black Holes in $f\left(  R\right)  $ Gravity}

\label{Sec:AGBHfG}

The black hole solution in $f\left(  R\right)  $ gravity can be found by
solving the gravitational equation $\left(  \ref{eq;EOM}\right)  $ plus some
possible matter equations for $g_{\mu\nu}$. However, it is quite complicated
and even impossible to find the analytical solutions in the general case.
Instead, one usually looks for the black hole solutions in $f\left(  R\right)
$ gravity with imposing the constant curvature condition. When $R=R_{0}$ which
is a constant, the trace of eqn. $\left(  \ref{eq;EOM}\right)  $ leads to%
\begin{equation}
2f^{\prime}\left(  R_{0}\right)  R_{0}-\left(  d+1\right)  f\left(
R_{0}\right)  =T^{m}, \label{eq:RT}%
\end{equation}
where $T^{m}$ is the trace of $T_{\mu\nu}^{m}$. Eqn. $\left(  \ref{eq:RT}%
\right)  $ implies that $T^{m}$ is also a constant. Moreover, it has been
shown in \cite{AGBHfG-Moon:2011hq} that $T^{m}=0$ to obtain the constant
curvature black hole solution in $f\left(  R\right)  $ gravity coupled to a
matter field. For example, one has $T^{m}=0$ in the cases of the vacuum and
Maxwell field with $d=3$. Moreover, when $R=R_{0}$, one has that%
\begin{equation}
\tilde{\nabla}_{\tilde{\mu}}\phi=\sqrt{\frac{2d}{d-1}}\tilde{\nabla}%
_{\tilde{\mu}}\ln f^{\prime}\left(  \chi\right)  =\sqrt{\frac{2d}{d-1}%
}\partial_{\tilde{\mu}}\ln f^{\prime}\left(  R_{0}\right)  =0.
\end{equation}
Hence, the $S_{\partial\mathcal{V}}^{\phi}=0$ for the black hole solution with
constant curvature.

\subsection{Schwarzschild-AdS Black Hole}

First we consider the static black hole solution with constant curvature in
vacuum, where $T_{\mu\nu}^{m}=0$. This black hole solution was obtained in
\cite{AGBHfG-delaCruzDombriz:2009et,AGBHfG-Moon:2011hq}:%
\begin{equation}
ds^{2}=-b\left(  r\right)  dt^{2}+\frac{dr^{2}}{b\left(  r\right)  }%
+r^{2}d\Sigma_{k,d-1}^{2},
\end{equation}
where%
\begin{equation}
b\left(  r\right)  =k-\frac{m}{r^{d-2}}+\frac{r^{2}}{L^{2}}\text{,}%
\end{equation}
the constant Ricci scalar $R_{0}\equiv-\frac{\left(  d+1\right)  d}{L^{2}}$,
and $d\Sigma_{k,d-1}^{2}$ is the line element of the $\left(  d-1\right)
$-dimensional hypersurface with constant scalar curvature $\left(  d-1\right)
\left(  d-2\right)  k$ with $k=\left\{  -1,0,1\right\}  $. The parameters $m$
is related to the ADM mass $M$ of the black hole by
\cite{AGBHfG-delaCruzDombriz:2009et,AGBHfG-Moon:2011hq}
\begin{equation}
M=f^{\prime}\left(  R_{0}\right)  \left(  d-1\right)  \Omega_{k,d-1}m,
\label{eq:mass}%
\end{equation}
where $\Omega_{k,d-1}$ denotes the dimensionless volume of $d\Sigma
_{k,d-1}^{2}$. For $k=0$ and $-1$, one needs to introduce an infrared
regulator to produce a finite value of $\Omega_{k,d-1}$. As usual, we let
$r_{+}$ denote the outer horizon position with $b\left(  r_{+}\right)  =0$.
The rescaled metric $\tilde{g}_{\tilde{\mu}\tilde{\nu}}$ is given by eqn.
$\left(  \ref{eq:rescale}\right)  $%
\begin{align}
\tilde{g}_{\tilde{\mu}\tilde{\nu}}d\tilde{x}^{\tilde{\mu}}d\tilde{x}%
^{\tilde{\nu}}  &  =f^{\prime}\left(  R_{0}\right)  ^{\frac{2}{d-1}}\left[
-b\left(  r\right)  dt^{2}+\frac{dr^{2}}{b\left(  r\right)  }+r^{2}%
d\Sigma_{k,d-1}^{2}\right] \nonumber\\
&  =-\tilde{b}\left(  \tilde{r}\right)  d\tilde{t}^{2}+\frac{d\tilde{r}^{2}%
}{\tilde{b}\left(  \tilde{r}\right)  }+\frac{\tilde{r}^{2}}{f^{\prime}\left(
R_{0}\right)  ^{\frac{2}{d-1}}}d\Sigma_{k,d-1}^{2},
\end{align}
where we define%
\begin{equation}
\tilde{r}=f^{\prime}\left(  R_{0}\right)  ^{\frac{2}{d-1}}r\text{, }\tilde
{t}=t,
\end{equation}
the rest coordinates of $\tilde{x}_{\mu}$ are the same as these of $x_{\mu}$,
and $\tilde{b}\left(  \tilde{r}\right)  =f^{\prime}\left(  R_{0}\right)
^{\frac{2}{d-1}}b\left(  r\right)  $. The outer horizon position is then given
by $\tilde{r}_{+}=f^{\prime}\left(  R_{0}\right)  ^{\frac{2}{d-1}}r_{+}$ such
that $\tilde{b}\left(  \tilde{r}_{+}\right)  =0$. As argued in
\cite{AGBHfG-delaCruzDombriz:2009et}, $f^{\prime}\left(  R_{0}\right)  $
should be positive otherwise the entropy of the black hole would be negative.
It also showed in \cite{AGBHfG-Pogosian:2007sw}, the effective Newton's
constant in $f\left(  R\right)  $ gravity being positive also required
$f^{\prime}\left(  R_{0}\right)  $ to be positive.

\begin{figure}[tb]
\begin{center}
\includegraphics[width=0.6\textwidth]{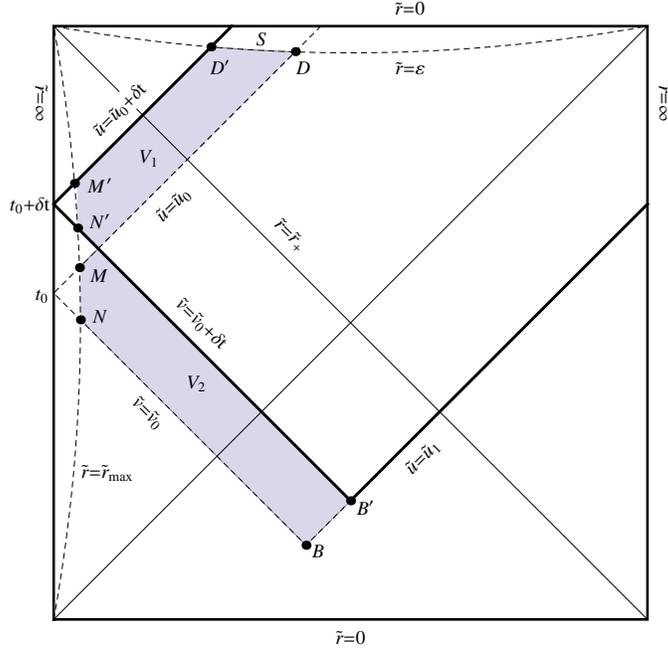}
\end{center}
\caption{Wheeler-deWitt patches of a Schwarzschild-AdS black hole in $f\left(
R\right)  $ gravity at $\tilde{t}_{L}=t_{0}$ and $\tilde{t}_{L}=t_{0}+\delta
t$. The lines $\tilde{r}=\tilde{r}_{\max}$ and $\tilde{r}=\varepsilon$ are the
cut-off surfaces.}%
\label{fig:prdiagram1}%
\end{figure}

We now use the methods in \cite{IN-Lehner:2016vdi} to calculate the change of
the action $\left(  \ref{eq:TotalS}\right)  $, $\delta S_{\text{WdW}%
}=S_{\text{WdW}}\left(  t_{0}+\delta t\right)  -S_{\text{WdW}}\left(
t_{0}\right)  $, of the Wheeler-DeWitt patch at late times. The Penrose
diagrams with the Wheeler-DeWitt patches at $\tilde{t}=t_{0}$ and
$t_{0}+\delta t$ are illustrated in FIG. $\ref{fig:prdiagram1}$. Fixing the
time on the right boundary, we only vary it on the left boundary. To regulate
a divergence near the boundary $\tilde{r}=\infty$, a surface of constant
$\tilde{r}=\tilde{r}_{\max}$ is introduced. We also introduce a spacelike
surface $\tilde{r}=\varepsilon$ near the future singularities and let
$\varepsilon\rightarrow0$ at the end of calculations. To calculate $\delta
S_{\text{WdW}}$, we introduce the null coordinates $\tilde{u}$ and $\tilde{v}%
$:%
\begin{align}
\tilde{u}  &  =\tilde{t}-\tilde{r}^{\ast}\nonumber\\
\tilde{v}  &  =\tilde{t}+\tilde{r}^{\ast},
\end{align}
where
\begin{equation}
\tilde{r}^{\ast}=\int\tilde{b}^{-1}\left(  \tilde{r}\right)  d\tilde{r}.
\end{equation}

Due to time translation, the joint contributions from $\mathcal{D}$ and
$\mathcal{D}^{\prime}$ are identical, and they therefore make no contribution
to $\delta S_{\text{WdW}}$. Similarly, the joint and surface contributions
from $\mathcal{MN}$ cancel against these from $\mathcal{M}^{\prime}%
\mathcal{N}^{\prime}$ on $\tilde{r}=\tilde{r}_{\max}$ in calculating $\delta
S_{\text{WdW}}$. Since $S_{\partial\mathcal{V}}^{\phi}=0$ and null surfaces
make no contribution to $\delta S_{\text{WdW}}$, eqn. $\left(  \ref{eq:TotalS}%
\right)  $ reduces to%
\begin{equation}
\delta S_{\text{WdW}}=S_{\mathcal{V}_{1}}-S_{\mathcal{V}_{2}}+2\int
_{\mathcal{S}}d^{d}\tilde{x}\sqrt{\left\vert \tilde{h}\right\vert }%
K+2\int_{\mathcal{B}^{\prime}}d^{d-1}\tilde{x}\sqrt{\tilde{\sigma}}%
a-2\int_{\mathcal{B}}d^{d-1}\tilde{x}\sqrt{\tilde{\sigma}}a.\label{eq:deltaSS}%
\end{equation}
Since the black hole solutions are on shell, the volume contribution can be
calculated by eqn. $\left(  \ref{eq:Sf(R)}\right)  $
\begin{align}
S_{\mathcal{V}} &  =f\left(  R_{0}\right)  \int_{\mathcal{V}}d^{d+1}x\sqrt
{-g}=f\left(  R_{0}\right)  f^{\prime}\left(  R_{0}\right)  ^{-\frac{d+1}%
{d-1}}\int_{\mathcal{V}}d^{d+1}\tilde{x}\sqrt{-\tilde{g}}\nonumber\\
&  =f\left(  R_{0}\right)  f^{\prime}\left(  R_{0}\right)  ^{-\frac{2d}{d-1}%
}\Omega_{k,d-1}\int_{\mathcal{V}}d\tilde{\omega}d\tilde{r}\tilde{r}%
^{d-1},\label{eq:vc}%
\end{align}
where $\tilde{\omega}=\left\{  \tilde{t},\tilde{u},\tilde{v}\right\}  $. The
region $\mathcal{V}_{1}$ is bounded by the null surfaces $\tilde{u}=\tilde
{u}_{0}$, $\tilde{u}=\tilde{u}_{0}+$ $\delta t$, $\tilde{v}=\tilde{v}_{0}+$
$\delta t$, the spacelike surface $\tilde{r}=\varepsilon$, and the timelike
surface $\tilde{r}=\tilde{r}_{\max}$. Using eqn. $\left(  \ref{eq:vc}\right)
$, we have that%
\begin{align}
S_{\mathcal{V}_{1}} &  =f\left(  R_{0}\right)  f^{\prime}\left(  R_{0}\right)
^{-\frac{2d}{d-1}}\Omega_{k,d-1}\int_{\tilde{u}_{0}}^{\tilde{u}_{0}+\delta
t}d\tilde{u}\int_{\varepsilon}^{\min\left\{  \tilde{r}_{\max},\rho\left(
\tilde{u}\right)  \right\}  }\tilde{r}^{d-1}d\tilde{r}\nonumber\\
&  =\frac{f\left(  R_{0}\right)  f^{\prime}\left(  R_{0}\right)  ^{-\frac
{2d}{d-1}}\Omega_{k,d-1}}{d}\int_{\tilde{u}_{0}}^{\tilde{u}_{0}+\delta
t}d\tilde{u}\tilde{r}^{d}|_{\tilde{r}=\min\left\{  \tilde{r}_{\max}%
,\rho\left(  \tilde{u}\right)  \right\}  }\text{,}%
\end{align}
where $\tilde{r}^{\ast}\left(  \rho\left(  \tilde{u}\right)  \right)  =\left(
\tilde{v}_{0}+\delta t-\tilde{u}\right)  /2$, and we neglect the
$\varepsilon^{d-1}$ term. Similarly for $\mathcal{V}_{2}$, we find that%
\begin{equation}
S_{\mathcal{V}2}=\frac{f\left(  R_{0}\right)  f^{\prime}\left(  R_{0}\right)
^{-\frac{2d}{d-1}}\Omega_{k,d-1}}{d}\int_{\tilde{v}_{0}}^{\tilde{v}_{0}+\delta
t}d\tilde{v}\tilde{r}^{d}|_{\rho_{1}\left(  \tilde{v}\right)  }^{\min\left\{
\tilde{r}_{\max},\rho_{0}\left(  \tilde{v}\right)  \right\}  },\label{eq:SV2S}%
\end{equation}
where $\tilde{r}^{\ast}\left(  \rho_{0/1}\left(  \tilde{v}\right)  \right)
=\left(  \tilde{v}-\tilde{u}_{0/1}\right)  /2$. Performing the change of
variables $\tilde{u}=\tilde{u}_{0}+\tilde{v}_{0}+\delta t-\tilde{v}$, one has
that%
\begin{equation}
\int_{\tilde{v}_{0}}^{\tilde{v}_{0}+\delta t}d\tilde{v}\tilde{r}^{d}%
|_{\tilde{r}=\min\left\{  \tilde{r}_{\max},\rho_{0}\left(  \tilde{v}\right)
\right\}  }=\int_{\tilde{u}_{0}}^{\tilde{u}_{0}+\delta t}d\tilde{u}\tilde
{r}^{d}|_{\tilde{r}=\min\left\{  \tilde{r}_{\max},\rho\left(  \tilde
{u}\right)  \right\}  },
\end{equation}
and hence%
\begin{equation}
S_{\mathcal{V}_{1}}-S_{\mathcal{V}_{2}}=\frac{f\left(  R_{0}\right)
f^{\prime}\left(  R_{0}\right)  ^{-\frac{2d}{d-1}}\Omega_{k,d-1}}{d}%
\int_{\tilde{v}_{0}}^{\tilde{v}_{0}+\delta t}d\tilde{v}\tilde{r}^{d}%
|_{r=\rho_{1}\left(  \tilde{v}\right)  },
\end{equation}
which shows that the portion of $\mathcal{V}_{1}$ below the future horizon
cancels against the portion of $\mathcal{V}_{2}$ above the past horizon. At
late times, one has that $\rho_{1}\left(  \tilde{v}\right)  \approx\tilde
{r}_{+}=f^{\prime}\left(  R_{0}\right)  ^{\frac{2}{d-1}}r_{+}$, and%
\begin{equation}
S_{\mathcal{V}_{1}}-S_{\mathcal{V}_{2}}=\frac{f\left(  R_{0}\right)
\Omega_{k,d-1}}{d}r_{+}^{d}\delta t\text{.}\label{eq:S1}%
\end{equation}

There is a timelike hypersurface at $\tilde{r}=\varepsilon$, with
outward-directed normal vectors from the region of interest. The normal vector
is%
\begin{equation}
\tilde{n}_{\mu}d\tilde{x}^{\mu}=\frac{-1}{\sqrt{-\tilde{b}\left(  \tilde
{r}\right)  }}d\tilde{r}\text{.}%
\end{equation}
The trace of extrinsic curvature is
\begin{equation}
K=\frac{1}{\tilde{r}^{d-1}}\partial_{\tilde{r}}\left(  \tilde{r}^{d-1}%
\sqrt{-\tilde{b}\left(  \tilde{r}\right)  }\right)  .
\end{equation}
Therefore, the surface contributions from $\tilde{r}=\varepsilon$ is%
\begin{equation}
2\int_{\mathcal{S}}d^{d}\tilde{x}\sqrt{\left\vert \tilde{h}\right\vert
}K=mdf^{\prime}\left(  R_{0}\right)  \Omega_{k,d-1}\delta t, \label{eq:S2}%
\end{equation}
where we use $\sqrt{\left\vert \tilde{h}\right\vert }=\sqrt{-\tilde{b}\left(
\tilde{r}\right)  }f^{\prime}\left(  R_{0}\right)  ^{-1}\tilde{r}^{d-1}$ and
$\tilde{b}\left(  \tilde{r}\right)  \sim-\frac{m}{\tilde{r}^{d-2}}f^{\prime
}\left(  R_{0}\right)  ^{2}$ for small $\tilde{r}$.

Following \cite{IN-Carmi:2016wjl}, the integrand $a$ in the joint terms of
eqn. $\left(  \ref{eq:deltaSS}\right)  $ is
\begin{align}
a  &  =\epsilon\ln\left\vert \mathbf{k}_{1}\cdot\mathbf{k}_{2}/2\right\vert
,\nonumber\\
\epsilon &  =-\text{sign}\left(  \mathbf{k}_{1}\cdot\mathbf{k}_{2}\right)
\text{sign}\left(  \mathbf{\hat{k}}\cdot\mathbf{k}_{2}\right)  ,
\end{align}
where for $\mathcal{B}$ and $\mathcal{B}^{\prime}$,
\begin{align}
\left(  \mathbf{k}_{1}\right)  _{\mu}  &  =-c_{1}\partial_{\tilde{\mu}}\left(
\tilde{t}+\tilde{r}^{\ast}\right)  ,\nonumber\\
\left(  \mathbf{k}_{2}\right)  _{\mu}  &  =c_{2}\partial_{\tilde{\mu}}\left(
\tilde{t}-\tilde{r}^{\ast}\right)  ,
\end{align}
and the auxiliary null vectors $\mathbf{\hat{k}}$ is the null vector
orthogonal to the joint and pointing outward from the boundary region.
Therefore, we find that%
\begin{equation}
2\int_{\mathcal{B}^{\prime}}d^{d-1}\tilde{x}\sqrt{\tilde{\sigma}}%
a-2\int_{\mathcal{B}}d^{d-1}\tilde{x}\sqrt{\tilde{\sigma}}a=2\Omega
_{k,d-1}\left[  \tilde{h}\left(  \tilde{r}_{\mathcal{B}^{\prime}}\right)
-\tilde{h}\left(  \tilde{r}_{\mathcal{B}}\right)  \right]  , \label{eq:2sigma}%
\end{equation}
where%
\begin{equation}
\tilde{h}\left(  \tilde{r}\right)  =f^{\prime}\left(  R_{0}\right)
^{-1}\tilde{r}^{d-1}\ln\left(  -\frac{\tilde{b}\left(  \tilde{r}\right)
}{c_{1}c_{2}}\right)  .
\end{equation}
At late times, we have that $\tilde{r}_{\mathcal{B}}\approx\tilde{r}_{+}$ and
\begin{equation}
\tilde{h}\left(  \tilde{r}_{\mathcal{B}^{\prime}}\right)  -\tilde{h}\left(
\tilde{r}_{\mathcal{B}}\right)  =\frac{\tilde{b}\left(  \tilde{r}\right)  }%
{2}\frac{d\tilde{h}\left(  \tilde{r}\right)  }{d\tilde{r}}|_{\tilde{r}%
=\tilde{r}_{B}}\delta t=\frac{f^{\prime}\left(  R_{0}\right)  ^{-1}\tilde
{r}^{d-1}}{2}\frac{d\tilde{b}\left(  \tilde{r}\right)  }{d\tilde{r}}%
|_{\tilde{r}=\tilde{r}_{+}}\delta t, \label{eq:h}%
\end{equation}
where we use $d\tilde{r}=\tilde{b}\left(  \tilde{r}\right)  \delta t/2$ on
$\tilde{u}=u_{1}$. Thus, this gives%
\begin{equation}
2\int_{\mathcal{B}^{\prime}}d^{d-1}\tilde{x}\sqrt{\tilde{\sigma}}%
a-2\int_{\mathcal{B}}d^{d-1}\tilde{x}\sqrt{\tilde{\sigma}}a=\Omega
_{k,d-1}f^{\prime}\left(  R_{0}\right)  r_{+}^{d-1}\left[  \left(  d-2\right)
\frac{m}{r_{+}^{d-1}}+\frac{2r_{+}}{L^{2}}\right]  \delta t, \label{eq:S3}%
\end{equation}
where we use $d\tilde{b}\left(  \tilde{r}\right)  /d\tilde{r}=db\left(
r\right)  /dr$. Combining eqns. $\left(  \ref{eq:S1}\right)  $, $\left(
\ref{eq:S2}\right)  $, and $\left(  \ref{eq:S3}\right)  $, we arrive at%
\begin{equation}
\delta S_{\text{WdW}}=2\left(  d-1\right)  f^{\prime}\left(  R_{0}\right)
\Omega_{d-1}m\delta t\text{,}%
\end{equation}
where we use eqn. $\left(  \ref{eq:RT}\right)  \,$\ with $T^{m}=0$. Since
$t=\tilde{t}$, eqn. $\left(  \ref{eq:mass}\right)  $ leads to
\begin{equation}
\frac{dS_{\text{WdW}}}{dt}=2M,
\end{equation}
which has the same form as for the SAdS black hole in the Einstein's gravity.

\subsection{Charged Black Hole}

To have a black hole solution with constant curvature, the trace of the
energy-momentum tensor of the matter filed must vanish
\cite{AGBHfG-Moon:2011hq}. It is obvious that the standard Maxwell
energy-momentum tensor is traceless in four dimensions but not in higher
dimensions. On the other hand, an extension of Maxwell action in $\left(
d+1\right)  $-dimensional spacetime that is traceless is the conformally
invariant Maxwell action \cite{AGBHfG-Hassaine:2007py}:%
\begin{equation}
S^{m}=-\int d^{d+1}x\sqrt{-g}\left(  F_{\mu\nu}F^{\mu\nu}\right)  ^{\left(
d+1\right)  /4}, \label{eq:IMax}%
\end{equation}
$F_{\mu\nu}=\partial_{\mu}A_{\nu}-\partial_{\nu}A_{\mu}$ is the
electromagnetic field tensor, and $A_{\mu}$ is the electromagnetic potential.
When $d=3$, the action $\left(  \ref{eq:IMax}\right)  $ recovers the standard
Maxwell action. The EOM obtained by varying the action $\left(
\ref{eq:action}\right)  $ with respect to $A_{\mu}$ is%
\begin{equation}
\partial_{\mu}\left(  \sqrt{-g}F^{\mu\nu}F^{\left(  d-1\right)  /4}\right)
=0,
\end{equation}
where $F=F^{\mu\nu}F_{\mu\nu}$. Together with the gravitational equation
$\left(  \ref{eq:RT}\right)  $, the black hole solution was given in
\cite{AGBHfG-Sheykhi:2012zz}:%
\begin{align}
ds^{2}  &  =-b\left(  r\right)  dt^{2}+\frac{dr^{2}}{b\left(  r\right)
}+r^{2}d\Sigma_{k,d-1}^{2},\nonumber\\
F_{tr}  &  =\frac{q}{r^{2}}, \label{eq:CBH}%
\end{align}
where
\begin{equation}
b\left(  r\right)  =k-\frac{m}{r^{d-2}}+\frac{q^{2}}{r^{d-1}}\frac{\left(
-2q^{2}\right)  ^{\left(  d-3\right)  /4}}{f^{\prime}\left(  R_{0}\right)
}+\frac{r^{2}}{L^{2}},
\end{equation}
and the constant Ricci scalar $R_{0}\equiv-\frac{\left(  d+1\right)  d}{L^{2}%
}$. To have a real solution, the dimensions $d+1$ must be multiples of four,
i.e., $d=3,7,\cdots$. The parameters $m$ and $q$ are related to the mass $M$
and charge $Q$ of the black hole by \cite{AGBHfG-Sheykhi:2012zz}%
\begin{align}
M  &  =f^{\prime}\left(  R_{0}\right)  \left(  d-1\right)  \Omega
_{k,d-1}m,\nonumber\\
Q  &  =\frac{\left(  d+1\right)  \left(  -2\right)  ^{\left(  d-3\right)
/4}q^{\left(  d-1\right)  /2}\Omega_{k,d-1}}{16\pi\sqrt{f^{\prime}\left(
R_{0}\right)  }}, \label{eq:MQ}%
\end{align}
and the electric potential $\Phi$ at the horizon radius $r_{\pm}$ is%
\begin{equation}
\Phi_{\pm}=16\pi\frac{q}{r_{\pm}}\sqrt{f^{\prime}\left(  R_{0}\right)  }.
\label{eq:Phi}%
\end{equation}
As argued before, one has $f^{\prime}\left(  R_{0}\right)  >0$ to obtain
physical solutions. The black hole solution $\left(  \ref{eq:CBH}\right)  $ is
similar to a Reissner-Nordstrom AdS black hole. Thus, this solution could have
two horizons at the outer radius $r_{+}$ and inner radius $r_{-}$,
respectively. When $r_{+}=r_{-}$, the black hole is extremal.

\begin{figure}[tb]
\begin{center}
\includegraphics[width=0.4\textwidth]{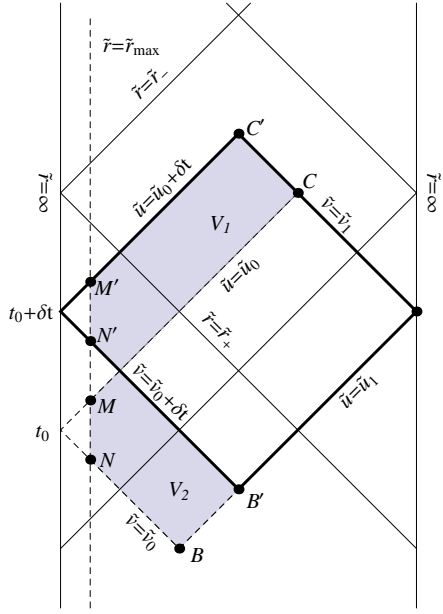}
\end{center}
\caption{Wheeler-deWitt patches of a charged black hole in $f\left(  R\right)
$ gravitiy at $\tilde{t}_{L}=t_{0}$ and $\tilde{t}_{L}=t_{0}+\delta t$. The
line $\tilde{r}=\tilde{r}_{\max}$ is the cut-off surface.}%
\label{fig:prdiagram2}%
\end{figure}

We now calculate the change $\delta S_{\text{WdW}}=S_{\text{WdW}}\left(
t_{0}+\delta t\right)  -$ $S_{\text{WdW}}\left(  t_{0}\right)  $ in the total
action $\left(  \ref{eq:TotalS}\right)  $ between the two WdW patches
displayed in FIG. $\ref{fig:prdiagram2}$. Taking time translation into
account, $\delta S_{\text{WdW}}$ reduces to
\begin{equation}
\delta S_{\text{WdW}}=S_{\mathcal{V}_{1}}-S_{\mathcal{V}_{2}}+2\int
_{\mathcal{B}^{\prime}}d^{d-1}\tilde{x}\sqrt{\tilde{\sigma}}a-2\int
_{\mathcal{B}}d^{d-1}\tilde{x}\sqrt{\tilde{\sigma}}a+2\int_{\mathcal{C}%
^{\prime}}d^{d-1}\tilde{x}\sqrt{\tilde{\sigma}}a-2\int_{\mathcal{C}}%
d^{d-1}\tilde{x}\sqrt{\tilde{\sigma}}a.
\end{equation}
For the black hole solution $\left(  \ref{eq:CBH}\right)  $, we find that the
volume contribution is%
\begin{equation}
S_{\mathcal{V}}=\Omega_{k,d-1}\int_{\mathcal{V}}d\tilde{\omega}F\left(
\tilde{r}\right)  ,\label{eq:SV}%
\end{equation}
where $\tilde{\omega}=\left\{  \tilde{u},\tilde{v}\right\}  $, and%
\begin{equation}
F\left(  \tilde{r}\right)  =-\frac{2f^{\prime}\left(  R_{0}\right)  }{L^{2}%
}\tilde{r}^{d}+f^{\prime}\left(  R_{0}\right)  ^{\frac{2\left(  d+1\right)
}{d-1}}\frac{\left(  -2q^{2}\right)  ^{\left(  d+1\right)  /4}}{\tilde{r}%
}\text{.}%
\end{equation}
For $\mathcal{V}_{1}$, its volume contribution is
\begin{equation}
S_{\mathcal{V}_{1}}=f^{\prime}\left(  R_{0}\right)  ^{-\frac{2}{d-1}}%
\Omega_{k,d-1}\int_{\tilde{u}_{0}}^{\tilde{u}_{0}+\delta t}d\tilde{u}F\left(
\tilde{r}\right)  |_{\tilde{\rho}_{1}\left(  \tilde{u}\right)  }^{\min\left\{
\tilde{r}_{\max},\rho\left(  \tilde{u}\right)  \right\}  }d\tilde{r},
\end{equation}
where $\tilde{r}^{\ast}\left(  \rho\left(  \tilde{u}\right)  \right)  =\left(
\tilde{v}_{0}+\delta t-\tilde{u}\right)  /2$ and $r^{\ast}\left(  \tilde{\rho
}_{1}\left(  \tilde{u}\right)  \right)  =\frac{\tilde{v}_{1}-\tilde{u}}{2}$.
Similarly, the volume contribution $S_{\mathcal{V}_{2}}\,$is
\begin{equation}
S_{\mathcal{V}2}=f^{\prime}\left(  R_{0}\right)  ^{-\frac{2}{d-1}}%
\Omega_{k,d-1}\int_{\tilde{v}_{0}}^{\tilde{v}_{0}+\delta t}d\tilde{v}F\left(
\tilde{r}\right)  |_{\rho_{1}\left(  \tilde{v}\right)  }^{\min\left\{
\tilde{r}_{\max},\rho_{0}\left(  \tilde{v}\right)  \right\}  },
\end{equation}
where $r^{\ast}\left(  \rho_{0/1}\left(  \tilde{v}\right)  \right)  =\left(
\tilde{v}-\tilde{u}_{0/1}\right)  /2$. Making the change of variables
$\tilde{u}=\tilde{u}_{0}+\tilde{v}_{0}+\delta t-\tilde{v}$, we find that at
late times,
\begin{align}
\delta S_{\mathcal{V}} &  =f^{\prime}\left(  R_{0}\right)  ^{-\frac{2}{d-1}%
}\Omega_{k,d-1}\left[  \int_{\tilde{v}_{0}}^{\tilde{v}_{0}+\delta t}d\tilde
{v}F\left(  \tilde{r}\right)  |_{r=\rho_{1}\left(  v\right)  }-\int_{\tilde
{u}_{0}}^{\tilde{u}_{0}+\delta t}d\tilde{u}F\left(  \tilde{r}\right)
|_{r=\tilde{\rho}_{1}\left(  \tilde{u}\right)  }d\tilde{r}\right]  \nonumber\\
&  =\Omega_{k,d-1}\left[  -\frac{2f^{\prime}\left(  R_{0}\right)  }{L^{2}%
}r^{d}+\frac{\left(  -2q^{2}\right)  ^{\left(  d+1\right)  /4}}{r}\right]
|_{r_{-}}^{r_{+}}\delta t,
\end{align}
where $r_{+}\,$and $r_{-}$ are the outer and inner horizon radius, respectively.

For the joint contributions from $\mathcal{B}$ and $\mathcal{B}^{\prime}$ at
late times, eqns. $\left(  \ref{eq:2sigma}\right)  $ and $\left(
\ref{eq:h}\right)  $ give%
\begin{gather}
2\int_{\mathcal{B}^{\prime}}d^{d-1}\tilde{x}\sqrt{\tilde{\sigma}}%
a-2\int_{\mathcal{B}}d^{d-1}\tilde{x}\sqrt{\tilde{\sigma}}a=\Omega
_{k,d-1}f^{\prime}\left(  R_{0}\right)  ^{-1}\tilde{r}^{d-1}\frac{d\tilde
{b}\left(  \tilde{r}\right)  }{d\tilde{r}}|_{\tilde{r}=\tilde{r}_{+}}\delta
t\nonumber\\
=\Omega_{k,d-1}f^{\prime}\left(  R_{0}\right)  \left[  \left(  d-2\right)
m-\frac{\left(  d-1\right)  q^{2}}{r_{+}}\frac{\left(  -2q^{2}\right)
^{\left(  d-3\right)  /4}}{f^{\prime}\left(  R_{0}\right)  }+\frac{2r_{+}^{d}%
}{L^{2}}\right]  \delta t. \label{eq:SBB}%
\end{gather}
Analogously to deriving eqn. $\left(  \ref{eq:SBB}\right)  $, we find that%
\begin{equation}
2\int_{\mathcal{C}^{\prime}}d^{d-1}\tilde{x}\sqrt{\tilde{\sigma}}%
a-2\int_{\mathcal{C}}d^{d-1}\tilde{x}\sqrt{\tilde{\sigma}}a=-\Omega
_{k,d-1}f^{\prime}\left(  R_{0}\right)  \left[  \left(  d-2\right)
m-\frac{\left(  d-1\right)  q^{2}}{r_{-}}\frac{\left(  -2q^{2}\right)
^{\left(  d-3\right)  /4}}{f^{\prime}\left(  R_{0}\right)  }+\frac{2r_{-}^{d}%
}{L^{2}}\right]  \delta t.
\end{equation}
Summing up all the contributions, we obtain%
\begin{equation}
\delta S_{\text{WdW}}=\Omega_{k,d-1}\left(  d+1\right)  \frac{\left(
-2\right)  ^{\left(  d-3\right)  /4}q^{\left(  d+1\right)  /2}}{r}|_{r_{+}%
}^{r-}\delta t.
\end{equation}
Using eqns. $\left(  \ref{eq:MQ}\right)  $ and $\left(  \ref{eq:Phi}\right)
$, we can write $dS/dt$ in terms of $Q$ and $\Phi_{\pm}$:%
\begin{equation}
\frac{dS_{\text{WdW}}}{dt}=Q\Phi_{-}-Q\Phi_{+}. \label{eq:ds/dtD}%
\end{equation}

\section{Discussion and Conclusion}

\label{Sec:Con}

In this paper, we used the approach proposed by Lehner \textit{et al.}
\cite{IN-Lehner:2016vdi} to calculate the change of the action of
Wheeler-DeWitt patches in $f\left(  R\right)  $ gravity. However, the method
proposed in \cite{IN-Lehner:2016vdi} only works for the Einstein--Hilbert
action. In section \ref{Sec:AfG}, we instead considered a (classically)
dynamically equivalent theory of $f\left(  R\right)  $ gravity, which was a
Brans-Dicke theory with Brans--Dicke parameter $\omega_{\text{BD}}=0$. After
transforming the Brans--Dicke action in the Jordan frame to the Einstein frame
by a conformal transformation, we showed that the action in Einstein frame was
the Einstein-Hilbert action plus the actions of the matter field and an
auxiliary field. In section \ref{Sec:AGBHfG}, the black hole solutions in
$f\left(  R\right)  $ gravity with constant curvature were discussed in the
cases of the vacuum and power-Maxwell field, respectively. In vacuum, the
black hole solution was a Schwarzschild-AdS black hole. Coupled to a
conformally invariant Maxwell field, the black hole solution was similar to a
higher dimensional Reissner-Nordstrom AdS black hole but only exist for
dimensions which are multiples of four. The results for the rate of the action
at late times are summarized as%
\begin{align}
\text{Schwarzschild-AdS black hole} &  \text{: }\frac{dS_{\text{WdW}}}%
{dt}=2M,\nonumber\\
\text{Charged black hole} &  \text{: }\frac{dS_{\text{WdW}}}{dt}=Q\Phi
_{-}-Q\Phi_{+},
\end{align}
where $M$ and $Q$ are the mass and charge of the black hole, respectively;
$\Phi_{\pm}$ are the electric potential evaluated at $r_{\pm}$, respectively.
It is noteworthy that these results in $f\left(  R\right)  $ gravity have the
same form as in Einstein's gravity.

Currently, there are two approaches to calculate the action of Wheeler-DeWitt
patches. In \cite{IN-Lehner:2016vdi}, contributions from null surfaces were
zero by choosing affine parameterizations while contributions from joints were
considered. On the other hand, no contributions from joints were considered in
\cite{IN-Brown:2015lvg}. However, contributions from spacelike/timelike
surface approaching the null surface were included there. Although these two
approaches seem quite different, it showed \cite{IN-Lehner:2016vdi} that they
yielded the same results for various black holes in Einstein's gravity. The
$dS_{\text{WdW}}/dt$ for a Schwarzschild-AdS black hole in $f\left(  R\right)
$ gravity was calculated in \cite{IN-Alishahiha:2017hwg} using the method of
\cite{IN-Brown:2015lvg}, and the result in \cite{IN-Alishahiha:2017hwg} is the
same as in our paper. It seems that both approaches may give the same result
in $f\left(  R\right)  $ gravity. Whether there is a reason for this
coincidence deserves further considerations.

In \cite{IN-Cai:2016xho}, the action growth of the Wheeler-DeWitt patches in
the cases of AdS-RN black holes, (charged) rotating BTZ black holes, AdS Kerr
black holes and (charged) Gauss-Bonnet black holes were calculated using the
method of \cite{IN-Brown:2015lvg}. It was found there that the results could
be written as%
\begin{equation}
\frac{dS_{\text{WdW}}}{dt}=\left(  M-\Omega J-Q\Phi\right)  _{+}-\left(
M-\Omega J-Q\Phi\right)  _{-}, \label{eq:ds/dtC}%
\end{equation}
where $\Omega$ and $\Phi$ are angular velocity and electrical potential of a
black hole, respectively; $J$ and $Q$ are the angular momentum and electric
charge of the black hole, respectively; the subscript $+/-$ stand for
evaluations at the outer and inner horizons, respectively. The same expression
for the results in \cite{IN-Pan:2016ecg} was also obtained in the case of
massive gravities. A general case was considered in \cite{CON-Huang:2016fks},
and it was proved that the action growth rate equals the difference of the
generalized enthalpy at the outer and inner horizons. In our paper, we showed
that the action growth for charged black holes in $f\left(  R\right)  $
gravity could also be rewritten as form of eqn. $\left(  \ref{eq:ds/dtC}%
\right)  $.

The Lloyd bound \cite{CON-Llo:2000} on the complexity growth for a holographic
state dual to a uncharged black hole reads \cite{IN-Brown:2015lvg}%
\begin{equation}
\mathcal{\dot{C}\leq}\frac{2M}{\pi\hbar},\label{eq:LN}%
\end{equation}
where $M$ is the mass of the black hole. We showed that by CA duality $\left(
\ref{eq:CA}\right)  $, the complexity growth for a Schwarzschild-AdS black
hole black hole in $f\left(  R\right)  $ gravity saturates the Lloyd bound
$\left(  \ref{eq:LN}\right)  $. It has been proved in \cite{CON-Yang:2016awy}
that under the strong energy condition of steady matter outside the Killing
horizon, black holes in CA duality obey the Lloyd bound. As noted in
\cite{IN-Brown:2015lvg}, the rate of the complexity of a neutral black hole is
faster than that of a charged black hole due to the existence of conserved
charges. This leads to that the Lloyd bound can be generalized for a charged
black hole with the charge $Q$:%
\begin{equation}
\mathcal{\dot{C}\leq}\frac{2}{\pi\hbar}\left[  \left(  M-Q\Phi\right)
_{+}-\left(  M-Q\Phi\right)  _{\text{gs}}\right]  ,\label{eq:LBoundQ}%
\end{equation}
where $\Phi$ is the potential at the horizon, and $\left(  M-Q\Phi\right)
_{+/\text{gs}}$ are $M-Q\Phi$ calculated at the outer horizon and in the
ground state, respectively. Treating the system as a grand canonical ensemble
implies that the ground state has the same potential as the black hole under
consideration. For charged black holes in $f\left(  R\right)  $ gravity, the
ground states are extremal black holes with $r_{+}=r_{-}$. Near extremality,
our previous paper \cite{IN-Tao:2017fsy} showed that the Lloyd bound $\left(
\ref{eq:LBoundQ}\right)  $ is usually violated for charged black holes. These
violations may have something to do with hair \cite{IN-Brown:2015lvg}.

For charged black holes $\left(  \ref{eq:CBH}\right)  $ with fixed potential
$\Phi_{+}=\Phi_{0}$ far away from the ground state, one has large black holes
with $r_{+}\gg L$. In this case, we find that%
\begin{gather}
2\left(  M-Q\Phi\right)  _{+}-2\left(  M-Q\Phi\right)  _{\text{gs}}%
=2f^{\prime}\left(  R_{0}\right)  \left(  d-1\right)  \Omega_{k,d-1}%
\frac{r_{+}^{d}}{L^{2}}\left[  1+\mathcal{O}\left(  \frac{L^{2}}{r_{+}^{2}%
}\right)  \right]  ,\\
\frac{dS_{\text{WdW}}}{dt}=Q\Phi_{-}-Q\Phi_{+}=f^{\prime}\left(  R_{0}\right)
\Omega_{k,d-1}\frac{\left(  d+1\right)  r_{+}^{d}}{L^{2}}\left[
1+\mathcal{O}\left(  \frac{L^{2}}{r_{+}^{2}}\right)  \right]  ,\nonumber
\end{gather}
which show that the Lloyd bound is satisfied but not saturated for charged
black holes with $d=7,11,\cdots$. However for $d=3$, we need to find higher
order terms to check whether the Lloyd bound is violated. When $d=3$, we
obtain%
\begin{gather}
2\left(  M-Q\Phi\right)  _{+}-2\left(  M-Q\Phi\right)  _{\text{gs}}=16\pi
f^{\prime}\left(  R_{0}\right)  \frac{r_{+}^{3}}{L^{2}}\left[  1+\frac{kL^{2}%
}{r_{+}^{2}}-\frac{1}{f^{\prime}\left(  R_{0}\right)  ^{2}}\left(  \frac
{\Phi_{0}}{16\pi}\right)  ^{4}\frac{L^{2}}{r_{+}^{2}}+\mathcal{O}\left(
\frac{L^{3}}{r_{+}^{3}}\right)  \right]  ,\\
\frac{dS_{\text{WdW}}}{dt}=Q\Phi_{-}-Q\Phi_{+}=16\pi f^{\prime}\left(
R_{0}\right)  \frac{r_{+}^{3}}{L^{2}}\left[  1+\frac{kL^{2}}{r_{+}^{2}%
}+\mathcal{O}\left(  \frac{L^{4}}{r_{+}^{4}}\right)  \right]  ,\nonumber
\end{gather}
which shows that the Lloyd bound is violated.

\begin{acknowledgments}
We are grateful to Houwen Wu and Zheng Sun for useful discussions. This work
is supported in part by NSFC (Grant No. 11005016, 11175039 and 11375121).
\end{acknowledgments}

\end{document}